# The SST Multi-G-Sample/s Switched Capacitor Array Waveform Recorder with Flexible Trigger and Picosecond-Level Timing Accuracy

Stuart A. Kleinfelder, *Senior Member, IEEE*, Edwin Chiem and Tarun Prakash

*Abstract*— The design and performance of a multi-G-sample/s fully-synchronous analog transient waveform recorder I.C. ("SST") with fast and flexible trigger capabilities is presented. Containing 4 channels of 256 samples per channel and fabricated in a 0.25 μm CMOS process, it has a 1.9V input range on a 2.5V supply, achieves 12 bits of dynamic range, and uses ~160 mW while operating at 2 G-samples/s and full trigger speeds. With a standard 50 Ohm input source, the SST's analog input bandwidth is ~1.3 GHz within about ±0.5 dB and reaches a -3 dB bandwidth of 1.5 GHz. The SST's internal sample clocks are generated synchronously via a shift register driven by an external LVDS oscillator, interleaved to double its speed (e.g., a 1 GHz clock yields 2 G-samples/s). It can operate over 6 orders of magnitude in sample rates (2 kHz to 2 GHz). Only three active control lines are necessary for operation: Reset, Start/Stop and Read-Clock. Each of the four channels integrates dual-threshold discrimination of signals with ~1 mV RMS resolution at >600 MHz bandwidth. Comparator results are directly available for simple threshold monitoring and rate control. The High and Low discrimination can also be AND'd over an adjustable window of time in order to exclusively trigger on bipolar impulsive signals. Trigger outputs can be CMOS or low-voltage differential signals, e.g. 1.2V CMOS or positive-ECL (0-0.8V) for low noise. After calibration, the imprecision of timing differences between channels falls in a range of 1.12-2.37 ps sigma at 2 G-samples/s.

## I. Introduction

ANALOG switched capacitor array ("SCA") waveform sampling and digitization circuits have seen wide use in applications that require large-scale, low-power transient signal acquisition. Early versions from 1988-1991 provided up to 13 bits of dynamic range, 100 MHz synchronously-clocked sample rates, and sequential or random-access sample addressing, e.g. [1]-[3]. The first G-sample/s version, published in 1992 [4], achieved 5 GHz sample rates by using an adjustable digital delay line to generate its high-speed sampling clocks. Various derivations have included parallel analog-to-digital conversion [5]-[7] and have been especially popular in the field of particle astrophysics, e.g. in the IceCube experiment [8] and elsewhere. Continued evolution led to on-chip phase-locked-loop clock generation and pattern-matching trigger circuitry, e.g. [9],[10]. Fast SCA's have been the subject of research and use by numerous other investigators, e.g. in [11]-[15], for their low power, wide dynamic range, the flexible features they afford, and their timing performance.

This paper describes a fully-synchronous multi-G-sample/s device, dubbed the "SST" for Synchronous Sampling plus Triggering [16]. Its principle advantages are that sample rates can span 6 orders of magnitude merely by changing the input clock speed, high dynamic range and timing precision, nearly flat Nyquist-rate bandwidth, and fast and sensitive trigger performance.

Fig. 1 shows a plot of the SST's layout, containing 4 channels of 256 samples per channel. Less than 20% of the rather small (~2.5 by 2.5 mm) chip includes the active sampling and trigger circuitry. For high performance, a 56-pin, 8 mm surface mount package was employed. The SST was designed for simplicity of operation, and only three active controls are required for operation: a Start/Stop pin that starts and stops sampling for convenient common-start or common-stop operation, a Reset pin, and a Read-Clock pin.

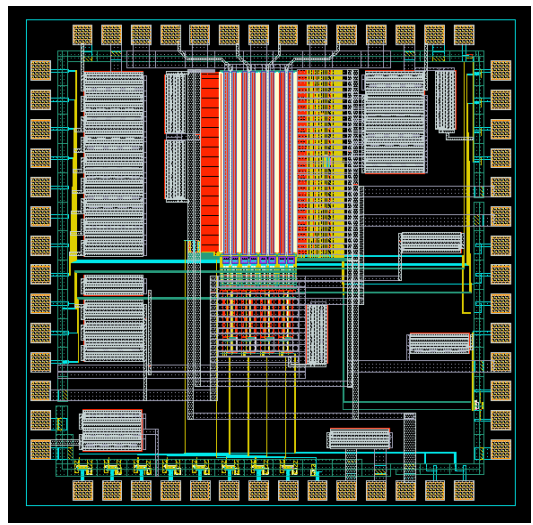

Fig. 1: A plot of the SST. The dimensions are ~2.5 by 2.5 mm.

The SST achieves nearly flat frequency response out to ~1.3 GHz and 1.5 GHz -3 dB bandwidth using a standard 50 Ohm

This work was supported by funding from the Office of Polar Programs and Physics Division of the U.S. National Science Foundation, grant awards ANT-08339133, NSF-0970175, and NSF-1126672, by NASA via grant 14-NIAC14B-0269, and by the Dept. of Physics and Astronomy, Uppsala University.
The authors are with the Department of Electrical Engineering and Computer Science at the University of California, Irvine CA, 92697, U.S.A. Email: stuartk@uci.edu.

input signal. It includes fast, flexible trigger capabilities with ~1 mV RMS sensitivity in response to pulses of 500 ps full-width at half-maximum. The trigger includes two thresholds per channel for bipolar triggering. A "window" feature can restrict triggering to bipolar pulses of a selected minimum frequency of interest (e.g., 100 MHz and above), helping to suppress triggers from thermal noise, etc. Individual comparator outputs can be supplied, either as the default mode or to aid in trigger calibration and threshold settings, or else low voltage differential trigger outputs can be generated. After calibration, the precision of timing measurements between channels falls in a range of 1.12-2.37 ps (sigma) at 2 G-samples/s.

## II. HIGH-SPEED SAMPLE CLOCK GENERATION

The SST's high speed sampling clocks are generated by a fast shift register containing a single "1" as a pointer. As shown in Fig. 2, the input clock is provided via an LVDS receiver which feeds a 2-phase clock generator. This drives a buffer tree that distributes the clock in a uniform fashion to the shift register, which consists of 128 master-slave flip-flops configured in a circular fashion. Both the master and slave sections of the shift register are used to generate sample clocks, and thus the sample rate is doubled by interleaving, e.g. a 1 GHz clock results in a 2 GHz sample rate.

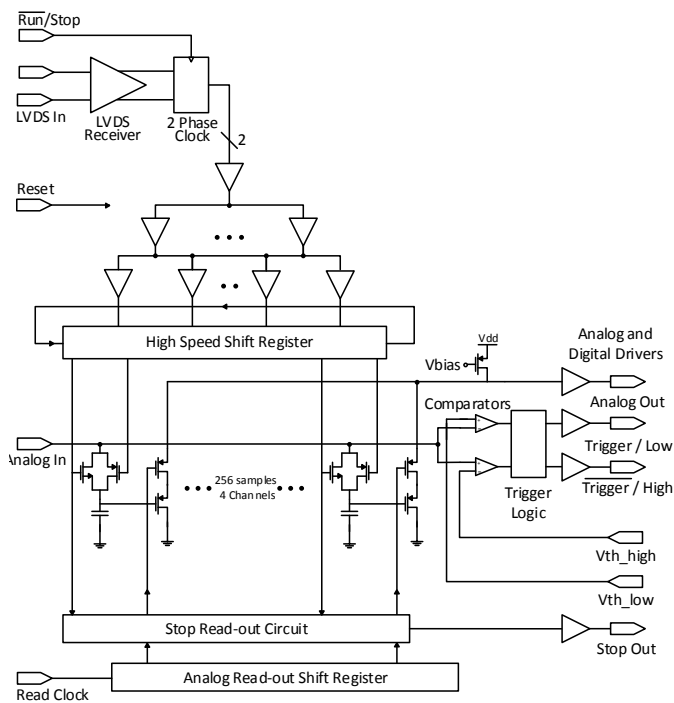

Fig. 2: Device overview showing clock features and one channel of four.

In many applications, the SST would be operated in a common-stop mode, endlessly sampling until it is stopped by a trigger, at which time the preceding samples contain the signal of interest. The position at which it is stopped may hence be random relative to the start, yet knowing its position remains important in order to delineate the beginning and end of the record. Therefore, the position of the sample clock "1" pointer (called the "Stop" in Fig. 2) at the moment it is stopped is read out in parallel with the analog samples.

Being implemented with simple, directly-clocked digital logic, the SST's sampling rate range has been measured to be over 6 orders of magnitude wide – from ~2 kHz, limited by the dynamic nature of the shift register – to at least 2 GHz, simply by changing its input clock frequency. This is a much wider range than asynchronous devices, which typically operate over only one order of magnitude in total frequency range. In addition, synchronous systems are not expected to suffer degraded timing precision at lower speeds, while this is a known weakness of asynchronous versions.

The SST can be started, stopped and synchronized immediately and with precision. In asynchronous systems that use a PLL, it is often necessary to keep much of the chip running constantly due to the time that PLL's require to establish lock, etc. The SST, in contrast, can be started or stopped within 1 ns, and samples at the start and stop of running are synchronized for high sample timing uniformity. Because a full high-speed clock is applied to the SST rather than a slow reference clock, this synchronization can also be performed externally if desired, and instant, flexible, on-the-fly control over the sample speed is attainable.

Fig. 3 shows a portion of a test board that was used for test results presented in this paper. There are 4 analog inputs at the left via SMA edge connectors, the SST is slightly left of center, and an FPGA that controls the SST's operation and readout is to the right. Bias-tee circuits are used between the SMA analog inputs and the SST in order to provide an AC-coupled arbitrary voltage offset at the SST's inputs. Termination of the analog inputs is by AC-coupled 50 Ohm resistors in order to reduce DC power consumption of the termination while using an input offset provided by the bias-tee's. A 1 GHz LVDS oscillator is used to clock the SST, which via interleaving then samples at 2 GHz. The rest of the board includes a microcontroller, switching regulators, power control and various communication components, etc. Thus, the measurements reported here include the non-ideal effects of a full, realistic system.

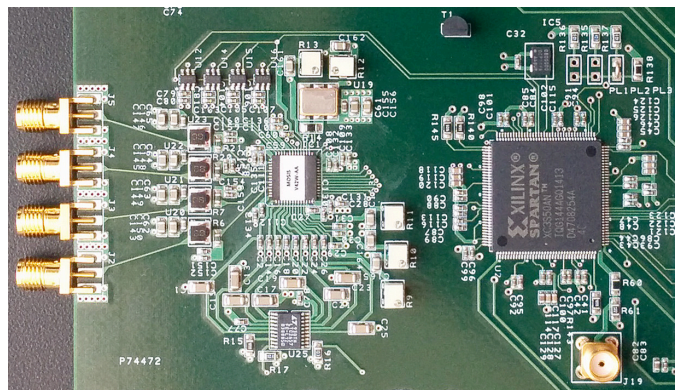

Fig. 3: Test board close-up showing 4 analog inputs (left edge, via SMA connectors), miniature bias-tee circuits to provide analog signal offsets, the SST chip (left of center), digital to analog conversion for threshold setting, analog to digital conversion, and an FPGA for managing the chip.

## III. ANALOG SAMPLING PERFORMANCE

The SST uses an economical 0.25 µm, 2.5 V process allowing a wide input range of ~1.9V to be achieved, leading to a dynamic range of ~12 bits. Sample capacitors are 80 fF each, implemented using a metal-insulator-metal technology. Complementary input switches are used for improved input range and more uniform bandwidth over that range, as well as reduced charge-injection issues. Two sample and holds are active at a time, with effort made to provide a "break before make" action. As seen in Fig. 2, above, a simple Active Pixel Sensor-type readout scheme was employed, using P-channel switched source-followers to match the sampled input range. Fig. 4 shows an example acquired waveform, that of a 100 MHz sine wave. Readers may note the very uniformly-spaced points noted from wave to wave that are indicative of low noise and excellent timing uniformity.

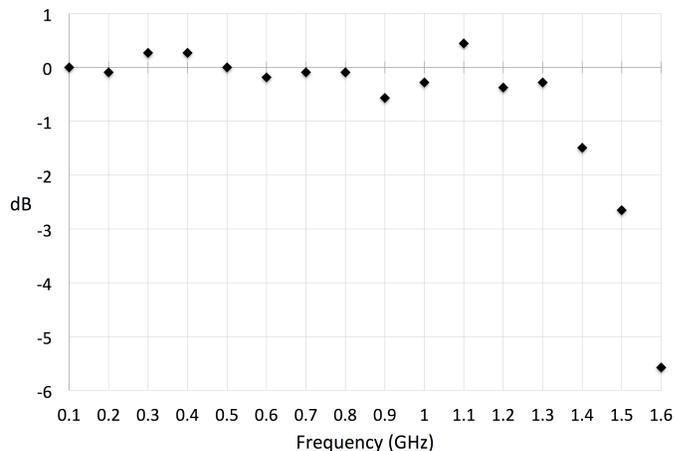

Fig. 5: Gain (magnitude) vs. frequency of the SST, normalized to 0 dB. The -3 dB point is just over 1.5 GHz.

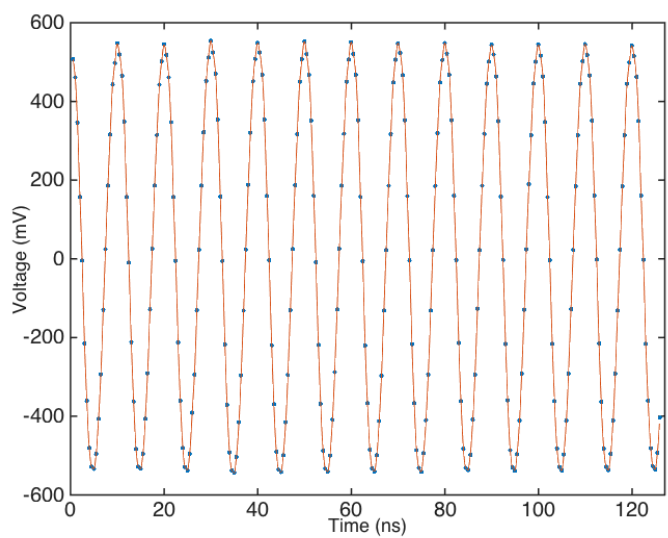

Fig. 4: SST readout, as digitized by the board shown in Fig. 3, of a 100 MHz sine wave sampled at 2 GHz after fixed pattern corrections.

The bandwidth was measured by applying sine waves of fixed amplitude but varying frequency from a signal generator with a standard 50 Ohm output impedance (Agilent N5181AEP-002), through the on-board bias-tee circuits, and AC-terminated at 50 Ohms on-board (no analog buffering was used). The SST recorded these sine waves at 2 G-samples/s, and the sampled analog output of the recorded sine waves was normalized to unity gain at low frequencies. As seen in Fig. 5, the resulting bandwidth was flat to about ±0.5 dB out to ~1.3 GHz, and the -3 dB bandwidth was found to be ~1.5 GHz.

The cell-to-cell fixed-pattern voltage offsets, which can be subtracted from the signal, were measured to be ~6.5 mV, RMS. The maximum input range of the SST is 0.05-1.95 V (1.9V total), but for good linearity a more typical range of operation is 0.1-1.8V. Over this 1.7 V range, the SST displays a worst-case integral non-linearity (deviation from a fit to the transfer-function's endpoints) of 17 mV, or 1%. Over the central 1V range, the worst-case integral non-linearity is 3 mV or 0.3%.

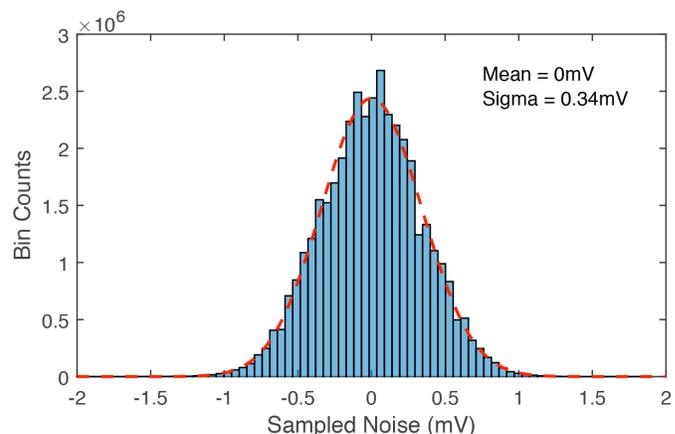

Fig. 6: Noise as measured from the board in Fig. 3, plotted from 10,000 events using all four channels, including 12-bit quantization and after pedestal subtraction, with an overlaid fitted Gaussian. The measured sigma is 0.34 mV.

A total system-level output noise sigma of 0.34 mV was measured (Fig. 6) after external digitization to 12 bits and pedestal subtraction, but with no other filtering or processing applied. Excluding the external 12-bit ADC's quantization noise, the standard deviation of the input-referred noise is 0.42 mV. In combination with the SST's 1.9V total input range, or its 1.7V linear (to 1%) input range, this yields a net dynamic range of 12 bits.

Finally, the leakage rate of the sample and holds was measured to be about 150 mV/s on average. The leakage is below the thermal noise of the chip at almost any target application sample rate, e.g. 1 MHz and above.

## IV. TRIGGER GENERATION AND PERFORMANCE

The SST incorporates high-performance real-time trigger circuitry that was designed to be sensitive to small, fast impulsive signals. One target application is the capture of brief radio-frequency chirps due to the Askaryan effect [17], currently of great interest in ultra-high-energy neutrino physics. The frequencies of interest in this application are generally in the 0.1-1 GHz range. In these and similar applications, a multi-level trigger system is often employed, and the SST has features to facilitate this.

## A. Comparator design and performance.

Each channel in the SST includes two fast comparators (Fig. 6) providing two thresholds per channel. These are typically used as a High and a Low threshold for bipolar signals (dual-threshold unipolar operation is also possible). Each pair of comparators observes its channel's analog input directly, with no gain, buffering or level-shifting before-hand. To achieve both high speed and gain, each comparator consists of 5 fully-differential stages plus a differential to single-ended conversion stage. An initial P-channel input stage is used to match the comparator's input range to that of the SST's analog input, followed by four stages using N-channel inputs. The differential stages have a gain of ~3.3 V/V each and a bandwidth of 1.2 GHz (-3 dB). The net low-frequency gain per comparator is over 2,000 V/V.

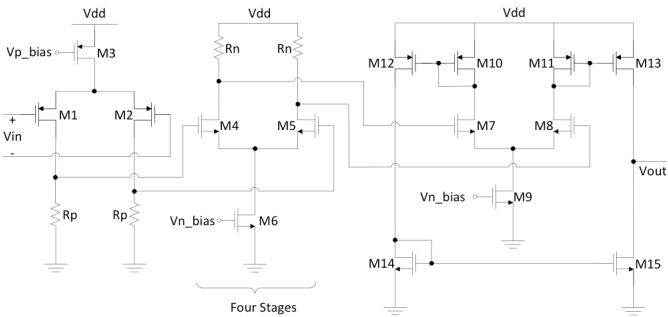

Fig. 6: Trigger comparator schematic. Note that four middle stages are used.

For the SST's target applications, and indeed for virtually any application, an ability to respond to small, fast pulses is important. Fig. 7 shows the SST triggering on an ~8 mV high, <500 ps full width at half maximum pulse from an Avtech model AVP-AV-1S-P-UCIA pulse generator (as measured by a 1 GHz bandwidth oscilloscope). A bias-tee was used to apply a DC offset, typically set to 0.9V when bipolar signals are expected. Successful triggering on this pulse would conservatively indicate a comparator bandwidth of over 600 MHz. The delay between the input pulse and the comparator output pulse, and after the trigger logic as described below, is dominated by cable and probe delays, and is actually about 2.6 ns. The "Level 2" output pulse width, as discussed in section IV-B, was arbitrarily set to ~17 ns. The post-pulse noise (about ±3 mV) seen in the upper input trace is substantially produced by the pulse generator itself, with some coupling (±2-3 mV) seen between the trigger output and the analog input. Note that the worst-case 2.5V single-ended trigger output mode was used rather than the SST's low-voltage differential trigger output mode, but the single-ended noise remained low enough that spurious triggers were not caused. The amplitude range of small pulses required to transition from 0% triggers to 100% triggers was found to be ~4 mV, indicating a trigger voltage sensitivity of ~1 mV RMS.

## B. Trigger decision logic.

For bipolar signals, combining two comparators with a High and a Low threshold can significantly reduce the rate of triggers due to random (e.g., thermal) noise. For significantly-enhanced efficiency, the high and low combination can be restricted to a range of time consistent with the frequencies of interest. In the SST, each channel's two comparators pass through pulse-stretching logic consisting of Set-Reset latches and digital delay lines (Fig. 8). The duration of the resulting stretched pulses provides the baseline for the channel's coincidence logic, which is a simple AND gate (an OR can be selected if desired). For example, if this Level-1 pulse-stretching is set to 5 ns (half of the wavelength of a 100 MHz bipolar signal), then a bipolar signal of 100 MHz or above may pass the AND'd trigger logic, but signals or noise below that frequency will tend to be suppressed.

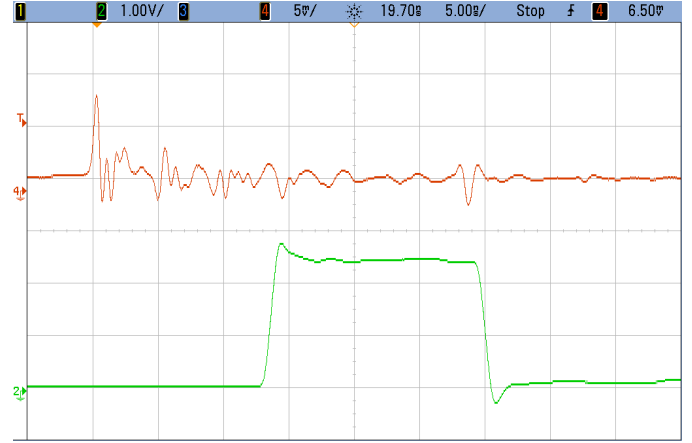

Fig. 7: Oscilloscope display (1 GHz bandwidth, 5 G-samples/s) of an ~8 mV input pulse of <500ps full-width at half maximum (upper trace), and the SST's trigger output response (lower trace). The top trace is 5 mV per division whereas the lower trace is 1 V per division.

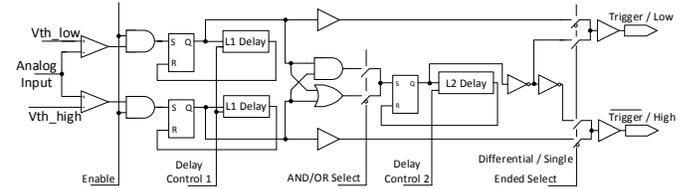

Fig. 8: Trigger logic including enable, pulse-stretching latches and delay lines, AND/OR selection and AND/OR pulse-stretching, and selection of dual single-ended or differential outputs.

The SST facilitates the simple tuning of comparator thresholds by allowing the AND/OR coincidence logic to be bypassed, thereby allowing each individual comparator to be monitored separately. External supervisory logic can thus easily record rates on individual comparators or form its own trigger on individual comparator results. This facility also allows the per-channel "High" and "Low" comparators to act as two different High or two different Low thresholds. In single-ended mode, a trigger output rate of over 200 MHz per channel is possible, and in AND mode, which passes through the Level-2 delay, a rate of over 100 MHz is possible.

## C. Trigger output voltage and delay.

Because the trigger must operate without causing significant interference, i.e. without causing cross-talk that may cause spurious triggers or corrupt the analog sampling, the SST can be configured to generate low-voltage differential trigger outputs. The outputs operate at their normal 2.5V CMOS logic levels, or else automatically reconfigure based on the desired voltage levels, with positive-ECL levels (0-0.8V) or 1.2V CMOS as the two targeted lower-voltage logic levels (see [16] for circuit details). In all results reported here, however, full 2.5V output signals were used. Using the SST's LVDS output mode would be expected to reduce cross-coupled noise by a factor of 3 or more.

Comparator delay is an often-quoted parameter in general applications, though for the target applications it is arguably much less important than trigger timing uniformity. Comparators typically respond more quickly when the voltage "overdrive" (the amount over threshold) is larger. Fig. 10 shows input pulse height vs. trigger delay after external cable delays are subtracted (but not the SST's own trigger logic delays such as the pulse stretching and output buffering circuitry). The input signal is the ~0.5 ns FWHM signal shown in Fig. 7, but with different amplitudes. As expected, the delay increases as the overdrive becomes smaller. The range is between ~2.22 ns to ~2.58 ns, or a span of ~0.36 ns over a voltage range of ~15mV to ~230 mV. This ~360 ps difference in delay is a modest percentage of the ~5ns Level 1 trigger window used in typical target applications, and is irrelevant compared to the total 128 ns endurance of the SST at 2 G-samples/s.

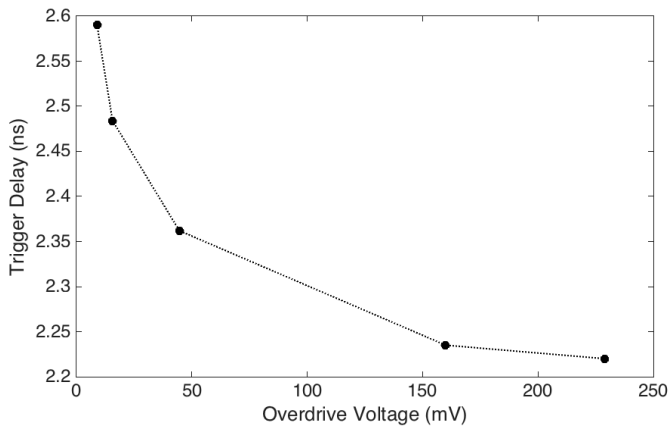

Fig. 9: Trigger delay (ns) vs. input pulse overdrive voltage. The delay includes the logic seen in Fig. 8 such as the pulse-stretching circuits and output buffers.

V. TIMING CALIBRATION BY SIMULATED ANNEALING

The following sub-sections describes the uncorrected timing fixed pattern noise (FPN) of the SST, a baseline measurement of the timing FPN, timing FPN obtained by simulated annealing calibration, and the SST's resulting timing performance within and between channels.

*A. Baseline fixed pattern timing noise.*

A fixed timing non-uniformity baseline was measured using a simple and commonly-used zero-crossing method (e.g. as in [14]). Sine waves were applied to every channel, asynchronous to the SST's clock, and the number of times a sine wave crossed its midpoint within each sample cell was independently counted. An "irrational" sine wave frequency is used; in this case 427.159 MHz at 1.6V from an Agilent N5181AEP-002 signal generator. The number of crossings accumulated per cell can then be scaled to calculate its width in time. The measured timing FPN is shown in Fig. 10.

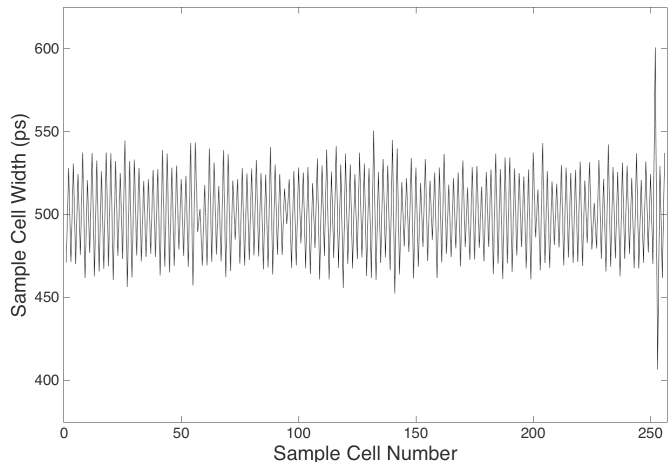

Fig. 10: Fixed pattern timing noise measured from an SST by using a common zero-crossing method. Typical randomly-distributed FPN is seen along with an odd/even interleaving effect. In addition, a substantial wrap-around effect is seen near the end of the record.

The SST exhibits two sources of systematic timing fixed pattern noise (FPN): an odd/even interleaving effect and a "wrap-around" effect. The odd/even effect is due to the fact that both clock phases are used generate sample clocks in order to double the sample rate relative to the clock rate, a form of FPN that is common in interleaved systems. In Fig 11 this amounts to about ±30 ps of timing non-uniformity, on average, on odd and even samples. The wrap-around from closing the shift register's loop also involves a timing non-uniformity on two samples: the sample generating the loop feedback and the immediately-subsequent sample, which given the clocked nature of the system must make up the timing difference. This amounts to about ±100 ps, or ~20% of the nominal 500 ps sample width. Look-ahead is used in the wrap-around circuit and hence the non-uniform samples are not at the very edges of the array. In retrospect, a few simple circuit changes could have substantially eliminated both sources of systematic timing differences, and a new version of this chip is now being fabricated that includes these changes.

In addition to systematic timing FPN, randomly-distributed yet fixed timing non-uniformities are present in all such systems, typically due to variations in transistor performance, capacitance values, etc., that emerge during the chip fabrication process. From Fig. 10, the SST was found to have about 7.3 ps, RMS, of randomly-distributed fixed timing variation (after ignoring systematic odd/even effects, etc.). On the other hand, fully-clocked systems such as the SST tend not to suffer from large-scale systematic timing changes across the sample and hold array that can result in linear, bowed or s-shaped timing non-uniformities.

The above timing imprecision, whether systematically or randomly-distributed, are fixed in nature and can be substantially cancelled by calibration such as that shown in

Fig. 10. In actual application, a system's timing performance will depend on many factors such as signal shape and amplitude, and covering all such possibilities is beyond the scope of this paper. Therefore, the following results may best be viewed as representative examples of the SST's performance potential.

### B. Intra-channel timing measurements.

The authors have investigated the use of simulated annealing to produce or optimize time-base FPN calibrations. The simulated-annealing heuristic typically begins with a model of the expected timing per sample. This could contain no timing FPN information at all (i.e., every sample is presumed to be exactly the same width in time; 500 ps each in these tests), or it could include starting FPN information from a circuit simulation (e.g., via SPICE) or from a measurement such as that shown in Fig. 10.

At each step of the simulated-annealing heuristic, a random modification of the time-base calibration is made by adding normally-distributed noise to correction values, producing a new tentative time-base calibration whose performance is then compared to the original. In this sub-section, that test is a calculation of the period durations of a set of acquired sine waves (in this case 1,000 waveforms of 333.333 MHz). Each period was calculated by simple two-point zero-crossing interpolations. The standard deviation of the resulting collection of period measurements over the entire data set was calculated, representing the over-all precision of the calibration. If the resulting sigma is lower (better) than the previous best, the winning FPN correction becomes the new FPN correction. As the process evolves, the "temperature" is gradually lowered, here by an adaptive reduction in the amount of added noise. The heuristic is stopped when the rate of incremental improvement, as judged by the temperature, falls below an arbitrary threshold.

Many versions of simulated annealing and similar heuristics may be employed. A common variation allows the acceptance of a percentage of solutions that are worse than the thus-far best solution in order to avoid becoming trapped in an undesirable corner of the solution space (as in the famous "hill climbing" problem), with the temperature reflecting the probability of accepting these "worse" solutions. This was tried as well, but no benefit was expected or observed.

Fig. 11 shows the evolution of the sigma of the sine wave periods during two example calibrations via simulated annealing. The starting condition for the top curve was completely uncalibrated (500 ps width per sample), yielding an initial 8.41 ps sigma in timing imprecision. The second began with the timing FPN as seen in Fig. 10, above, with an initial 2.35 ps sigma. Plotted on a semi-log scale, the heuristic was terminated after 500 adaptive temperature reduction iterations of 100 randomized-trial iterations each.

Both simulations completed with essentially identical sigmas of 2.1 and 2.16 ps. This was limited by the finite speed of sample acquisition, the simple linear interpolation used to calculate zero-crossing positions (which introduces a degree of imprecision when operating on sine waves), as well as the noise and jitter of the signal generator producing the sine waves, jitter in the SST's external clock, the thermal noise of sampling translated into the temporal domain, plus the inevitable temporal jitter introduced by the SST itself due to thermal noise in its clocking-related components, etc.

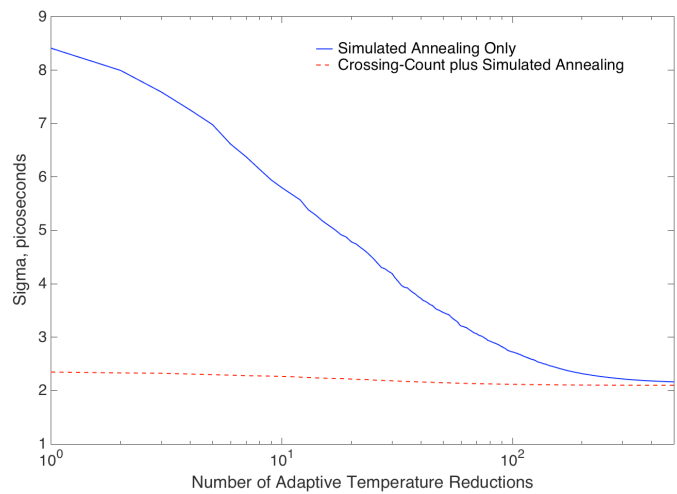

Fig. 11: Progression of a simulated annealing heuristic run in reducing the sigma of the measurement of the period of recorded sine waves. A log scale is used for the X axis.

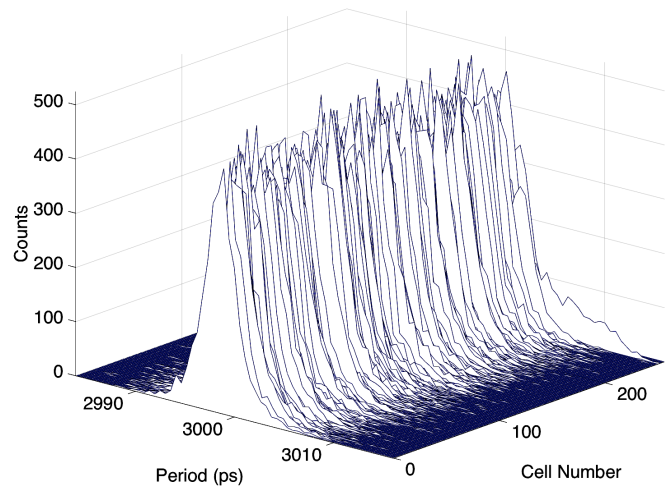

Fig. 12: Multiple-histogram plot of the measurement in ps of the period of a 1/3 GHz sine wave vs. sample cell number, after refinement by simulated annealing. The sigma of the histogram formed from all 256 samples is 2.1 ps.

Despite starting with no timing FPN information except for a presumption that each sample was 500 ps wide, the simulated annealing heuristic bested the measured calibration after about 50 cycles or ~10% of its running time, and by using only 1,000 sine waves vs. 2,000,000 in the Fig. 10 case.

Fig. 12 shows the collected period data after correction, organized as an array of histograms. It shows the measured periods of the 333 MHz sine waves vs. sample number (where each period started), computed from 20,000 waveforms. Cells with anomalous distributions can be seen near the end of the array that are associated with the high speed sample pointer's wrap-around. The uniformity across samples was improved to 2.1 ps (sigma).

## C. Inter-channel timing measurements

Tests were made in which the timing between signals on two separate channels was examined. The authors' test board is intended for bipolar signals and includes sub-miniature AC-coupled bias-tee circuits with relatively short time-constants. Therefore, a single 10 ns-wide bipolar sinusoidal pulse was used as a test stimulus, produced by an Agilent 81160A arbitrary waveform generator. Two copies of this waveform, one delayed with respect to the other by a Mini-Circuits model ZFRSC-42-S+ power splitter plus cables of varying lengths and quality, were applied to two channels.

Notably, the SST's own on-chip trigger circuitry was used in the data collection process: The SST was run in a common-stop mode, endlessly sampling and waiting until two pulses were seen on any two channels within a specified time window. When such a trigger was generated, the chip was stopped and read-out. Hence, these measurements include any noise or other non-idealities generated by the SST and the system board's triggering hardware. The arrival of the pulses were asynchronous with respect to the SST's operation, and thus the pulses fell on all samples in a random fashion.

After data collection, basic per-sample baseline voltage FPN and gain corrections were made. Several different timing FPN calibrations were then applied, including: (A) No calibration; each sample was assumed to be exactly 500 ps wide, (B) calibration by the zero-crossing method as discussed in Section V-*A*, and (C) calibration by simulated annealing, using 500 ps per sample (no calibration) as the starting condition.

Only one calibration per chip was produced in each case, rather than one per channel. This was sufficient due to the SST's high channel-to-channel timing uniformity. The "worst case" no-calibration scenario, (A), may be adequate for many applications, with the obvious advantage of requiring no calibration or computational effort. The common zero-crossing method, (B), was achieved by counting and then scaling the number of times asynchronous sine waves crossed zero within each sample cell (see Section V-*A*). This is a simple process but requires the acquisition of a large amount of data to achieve high accuracy (2 M events were used).

Calibration solely by simulated annealing, (C), started with an initial condition assuming that each sample was exactly 500 ps wide (no calibration). The heuristic's task was to minimize the standard deviation of the inter-channel cable delays using five representative delay times spanning 0 ns to 61 ns. Each delay was calculated by using a simple 2-point linear interpolation of the zero crossing of each bipolar pulse. This form of interpolation is subject to higher imprecision than, for example, fitting a curve to the pulse, but it is computationally-efficient.

Fig. 13 shows the progression of simulated annealing by plotting the sigma of each of the five delays as the temperature is adaptively lowered. In this example, no constraints were placed on the individual delays, and the sigmas of the individual delays were free to increase if that resulted in a lower average sigma. Indeed, the 4 ns delay curve moved modestly upward near the end of the progression.

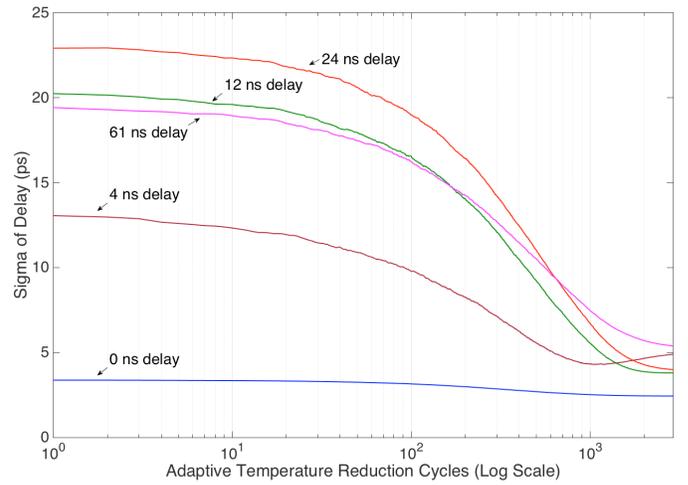

Fig. 13: Progression of the simulated annealing heuristic while optimizing five different delays between channels, starting with no calibration (500 ps per sample period) and using 2-point linear interpolation to find the zero-crossing of each bipolar pulse. The sigma of the delay over 1,000 waveform pairs is plotted for each delay time, with the X-axis on a log scale.

Fig. 14 shows the FPN resulting solely from applying the simulated annealing heuristic achieved in Fig. 13. Emerging solely from a uniform 500 ps per sample starting condition, it strongly resembled but yielded slightly larger deviations from the chip's odd/even and wrap-around effects than the zero-crossing measurement seen in Fig. 10, while offering slightly better performance.

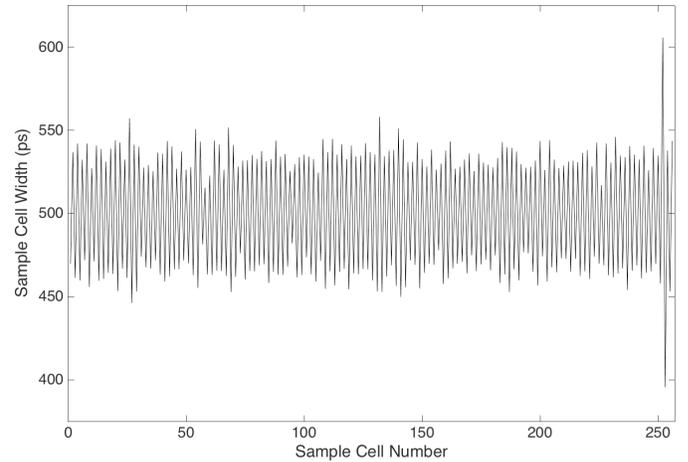

Fig. 14: Resulting timing FPN using simulated annealing only, with the starting condition of 500 ps per sample cell width (no calibration). This can be compared with Fig. 11.

Using simple two-point linear interpolations of the zero crossing of the bipolar pulses, a technique that is fast but not very precise, results in Fig. 15. Superior methods of measuring timing delays between channels would fit a function to the pulses or calculate the cross-correlation of each pair of waveforms, e.g. in [15] and [18]-[20]. Fig. 16 thus shows the result via the interpolation of each entire waveform to 100 fs increments followed by the calculation of the cross-correlation between each of the two full acquired records.

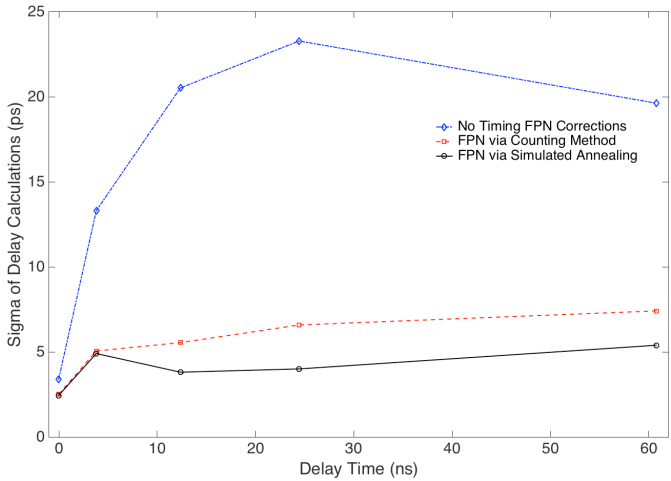

Fig. 15: Comparison of three timing FPN calibrations vs. the delay between pulses calculated by 2-point linear interpolation, including no timing FPN correction (top curve), an FPN calibration by the zero-crossing method (middle curve) and FPN correction by simulated annealing (bottom curve).

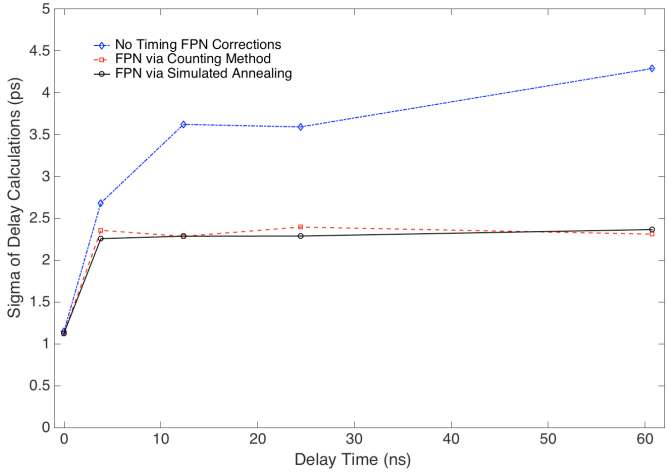

Fig. 16: Comparison of three timing FPN calibrations vs. the delay between pulses calculated by cross-correlation, including no timing FPN correction (top curve), an FPN calibration by the zero-crossing method and FPN correction by simulated annealing (bottom two curves).

With no time-base correction at all, the SST recorded under 4.3 ps standard deviation for all measured delays, a result that would suffice in a large number of applications. With time-base correction, the SST shows its best cross-channel timing precision (sigma of 1.12 ps) with zero time difference between channels. This is expected since non-idealities such as cross-talk, cable attenuation and dispersion, timing jitter due to the analog input signal generator or the on-board LVDS oscillator, etc., should be well-correlated and thus substantially cancel during the delay subtraction. Noting that the zero-delay case has the same temporal noise value regardless of calibration, including with no calibration at all, there is some reason to believe that it represents the net non-fixed temporal noise from all sources, such as jitter in the LVDS oscillator or in the analog input signal, etc.

Beyond 0 ns, the corrected profiles are almost flat, demonstrating the good timing uniformity that is the first-order expectation for a fully-synchronous system like the SST. A good match was also apparent between the crossing-count method and the simulated annealing results, which may indicate that limits due to thermal noise, etc., have been met.

A worst-case timing imprecision of 2.37 ps (sigma) was achieved between channels at any measured delay. This is a close match to the intra-channel precision found in section V-B. The flatness of the result and its apparent match to a fully-independent measurement again lends credence to the notion that fundamental limits have been reached, at least given the current system board and LVDS oscillator, the analog pulse generator used, etc. Finally, if one assumes the independence of the measurements of the arrival times of the two pulses on two channels, the worst-case single-measurement sigma would be $\sqrt{2}$ less, at 1.68 ps.

### D. Cautions about context and applicability.

Simulated annealing is a "black box" heuristic. It optimizes through random changes, and hence its results may not reflect underlying physical processes, and they may not necessarily resemble outcomes (e.g., FPN patterns) of other methods despite producing equivalent or better end results. Caution is advised when switching contexts, i.e. calibrating using one sort of measurement but using the calibration in another. It's best-suited for use under real-life application conditions by training it on a broad spectrum of real events.

### VI. SUMMARY, FUTURE WORK AND CONCLUSION.

A fully synchronous 2 GHz analog transient waveform recording and triggering I.C., the "SST," containing 4 channels of 256 cells per channel with fast and flexible on-chip trigger circuitry and ps-level timing accuracy was described. The SST was fabricated in a 0.25μm, 2.5V process to preserve a wide input range (0-1.9V), low noise (~0.42 mV input-referred RMS at room temperature) and low signal leakage (~150 mV/s) for high dynamic range (12 bits). Optimized design and packaging yielded a -3 dB bandwidth of ~1.5 GHz using a standard 50-Ohm signal source. Power consumption is ~160 mW at 2 GHz including full trigger power. The SST includes a dual-threshold coincidence trigger per channel that operates with ~1 mV RMS resolution with at least 600 MHz bandwidth. Example inter-channel timing correlation resolution was found to range between 1.12-2.37 ps RMS over a 0 to 61 ps timing difference time interval range. A performance summary is given in Table I.

Work in progress includes the design of an SST variation (in fabrication at the time of writing) that aims at the reduction or elimination of systematic fixed pattern timing non-uniformity, e.g. from interleaving and from closing the sample clock loop. The former is achieved by using two tightly-interleaved sample clock shift registers rather than both stages of a single master-slave shift register. The latter is achieved by balancing the loads on every shift register regardless of whether it is used as the feedback stage or not. Randomly-distributed fixed timing variations are inevitable, but these can be reduced as well with layout optimization, e.g. by avoiding the use of minimum-sized transistors at critical circuitry. Efforts are also aimed at reducing thermal timing noise by the reduction in the number of stages of electronic elements (i.e., from the clock inputs to the sample and hold clock signals) over which this can accumulate, and again by optimizing layout.

The SST has been successfully deployed in Antarctica in pilot ARIANNA ultra-high-energy neutrino detector systems [21]. Its flexibility and performance makes it attractive in a very wide range of applications.

Table I: SST Figures of Merit

| Parameter | Value |
|---|---|
| Technology: | 0.25 µm CMOS, 2.5V |
| Number of channels: | 4 |
| Samples per channel: | 256 |
| Chip size: | 2.5 by 2.5 mm |
| Package size: | 8 mm by 8 mm, 56 pins |
| Input clock (typical): | 1 GHz LVDS |
| Sample rate (typical): | 2 GHz |
| Minimum sample rate: | ~ 2 kHz |
| Analog bandwidth: | ~ 1.5 GHz, -3dB |
| Max power per chip with trigger: | ~ 160 mW at 2 GHz |
| Maximum analog input range: | ~ 0.05-1.95 V |
| Input-referred temporal noise: | ~ 0.42 mV, RMS |
| Dynamic range: | ~ 12 bits, RMS |
| Fixed pattern noise: | ~ 6.5 mV, RMS |
| Integral non-linearity (0.1-1.7V): | ~ 1 % over 1.6 V |
| Integral non-linearity (0.4-1.4V): | ~ 0.3 % over 1V |
| Crosstalk from a 1.6V sine wave: | <1 % at 300 MHz |
| Leakage rate: | ~ 0.15 V/s |
| Trigger comparators per channel: | 2 (typ. High and Low) |
| Trigger sensitivity: | < 1 mV, RMS |
| Trigger bandwidth: | > 600 MHz |
| Trigger functions per channel: | 2, AND/OR, windowed |
| Trigger output modalities: | Differential/Individual |
| Trigger output voltage: | 0.8, 1.2 or 2.5V CMOS |
| Trigger delay variation (~max.): | ~ 360 ps |
| Intra-channel timing resolution: | ~ 2.1 ps sigma |
| Uncorrected inter-channel res.: | ~ 1.15 - 4.3 ps sigma |
| Inter-channel timing resolution: | ~ 1.12 - 2.37 ps sigma |
| Inter-channel resolution ÷√2: | ~ 0.79 - 1.68 ps sigma |

ACKNOWLEDGMENTS

The authors thank Wei Cai for her assistance designing the SST's LVDS receiver, and Anirban Samanta for the design of the board seen in Fig. 3. The authors also thank Gary Varner for an interesting discussion, and especially Steven Barwick, Corey Reed, James Walker, Chris Persichilli and other members of the ARIANNA collaboration for their numerous helpful conversations and for their efforts on the ARIANNA system that utilizes the SST chip.